\documentstyle[preprint,aps,floats,psfig]{revtex}
\begin{document}
\newcommand{\beq}{\begin{equation}}
\newcommand{\eeq}{\end{equation}}
\draft
\title  {Resonant propagation of fluxons
in corner junctions with triplet pairing symmetry}
\author{ N. Stefanakis}
\address{CNRS - CRTBT, 25 Avenue des Martyrs
        BP 166-38042 Grenoble c\'edex 9, France}
\date{\today}
\maketitle

\begin{abstract}

We present the numerical solutions for the $I-V$ characteristics, 
and describe the motion of fluxons in a frustrated 
Josephson junction made of an unconventional triplet superconductor and an $s$-wave superconductor. In the inline geometry, and long length limit 
the moving integer fluxon interacts with the bound fractional fluxon but 
is not able to change its position or polarity. 
We observe different modes of multifluxon propagation.
In the small length limit the moving fluxon is a combination of the 
stable solutions that exist in the static case 
and additional steps are introduced in the $I-V$ 
diagram. 
\end{abstract}
\newpage

\section{Introduction}

The order parameter symmetry in the recently discovered superconductor Sr$_2$RuO$_4$ \cite{maeno} is an active area of current work. 
The Knight shift shows no change when passing through the superconducting state and this indicates that the pairing state is triplet \cite{ishida}. 
The magnetic field is spontaneously induced as shown by $\mu SR$ experiment, 
and this is a clear indication that 
the time-reversal symmetry is broken \cite{luke}. 
Inelastic-neutron-scattering measurements on single crystals of Sr$_2$RuO$_4$
show that the pairing state is highly anisotropic \cite{servant}.
Also specific heat measurements support the scenario of line nodes within the 
gap as in high-$T_c$ cuprate superconductors \cite{nishizaki}.

In Josephson junctions made of unconventional $d$-wave superconductors, the Josephson critical current
can be negative depending on the 
orientation angles of the crystallographic axes with respect to the junction interface. A negative critical current can be thought of as a phase 
shift of $\pi$ at the junction interface.
A half fluxon (antifluxon) is trapped at a $0 - \pi$ junction. 
It's existence is confirmed experimentally by measurements 
of the critical current versus the magnetic flux in corner junctions 
or corner SQUID where a dip appears in the critical current 
for magnetic field equal to zero \cite{woll}. The half magnetic flux 
quantum, ($\pi$ fluxon) 
has been directly observed using a scanning superconducting 
quantum interference device microscope
in tricrystal frustrated junctions
in superconducting YBa$_2$Cu$_3$O$_{7-\delta}$ \cite{tsuei}.

The static properties of one dimensional frustrated Josephson junctions 
between singlet and triplet superconductors have been presented for 
different nodal and nodeless pairing states \cite{stefan4}. 
The critical current and the spontaneous flux show a 
characteristic modulation with the junction orientation, which can be 
tested by experiment. 

The Josephson junction between $s$-wave superconductors supports 
modes of resonant propagation of fluxons \cite{fulton}. In the plot of the 
current-voltage ($I-V$) characteristics these modes appear as 
near-constant voltage branches known as zero field steps (ZFS)
\cite{chen,lomdahl}.
They occur in the absence of any external field. 
The ZFS appear at integer multiplies of $V_1=h c_S /2eL$, where 
$c_S$ is the velocity of the 
electromagnetic waves in the junction, and $L$ is the junction 
length.
The moving soliton 
is accompanied by a voltage pulse which can be detected at the 
junction's edges.

In Josephson junctions made of unconventional $d$-wave superconductors
the bound half-fluxon (-antifluxon) reverses its sign 
and emits an integer fluxon (antifluxon)
when biased by an external current \cite{kuklov,kato,stefan3}.
In this work we study the dynamic properties of a frustrated junction,
between singlet and triplet superconductors 
and calculate the $I-V$ characteristics. 
For the triplet superconductor Sr$_2$RuO$_4$ we shall assume
two possible pairing states of two
dimensional order parameter,
breaking the time reversal symmetry.
The first one is the nodeless $p$-wave order parameter 
with $E_u$ symmetry \cite{rice}.
The other one is the $f$-wave state
proposed by Hasegawa {\it et al.},
having $B_{1g}\times E_u$ symmetry \cite{hasegawa}. 

We study both the long and short length junction limit.
In the long length junction limit, 
the external current cannot move the fractional fluxon (antifluxon) 
$ff(faf)$ which is confined at $x=0$
and the ZES exist
at integer values of the
dc voltage $V_1$.
In the short length junction limit,
additional modes in the $I-V$ diagram
are introduced at half integer values of the
dc voltage $V_1$ 
and this can be used to distinguish the possible 
pairing states in the junction.

In the following we present the theoretical model for 
the corner junction in Sec. II. 
We present the results for the 
large junction limit in Sec. III.
The case of the shorter junction
is presented in Sec. IV
and finish with the conclusions.

\section{Corner junction model} 

\begin{figure}
\centerline{\psfig{figure=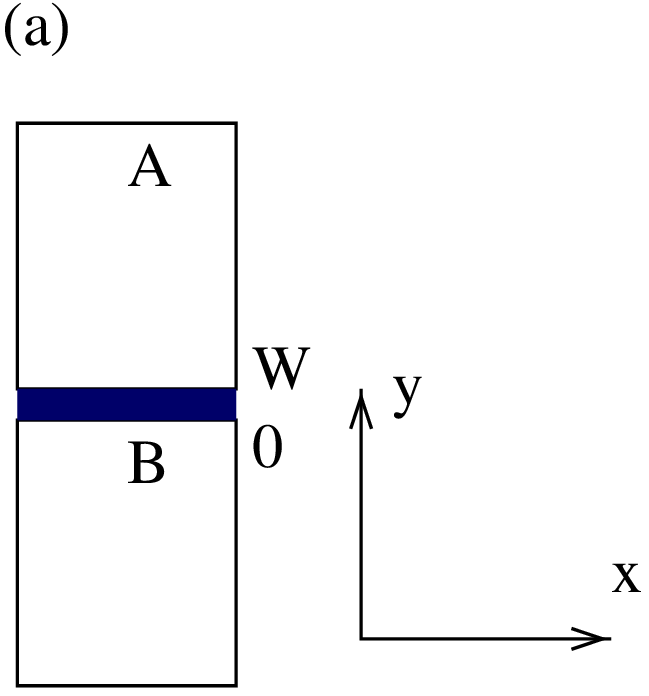,width=6cm,angle=-90}}
\centerline{\psfig{figure=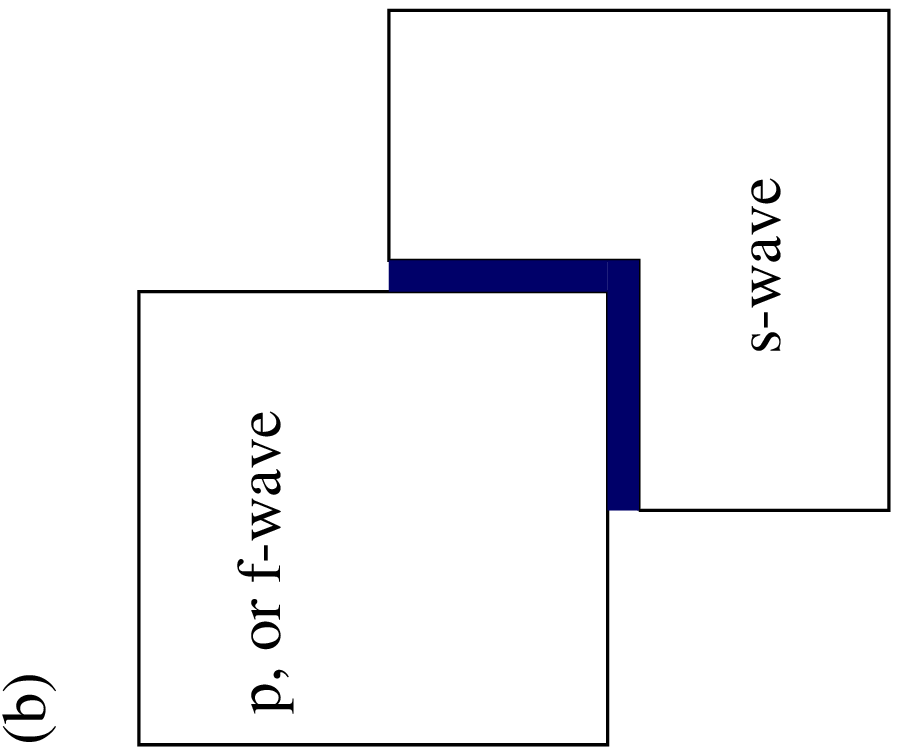,width=6cm,angle=-90}}
\caption{(a) View of the junction between a two component
triplet superconductor A
and a singlet superconductor B. The shaded region marks the 
interface which extends from $y=0$ to $y=W$.
(b) The corner junction geometry.
}
\label{figdyn.fig}
\end{figure}

We consider the junction shown in Fig. 1(a) between a triplet 
superconductor $A$ 
with a two component order parameter and an $s$-wave superconductor $B$.
The Ginzburg - Landau (GL) equations for isotropic p-wave superconductors
have been derived microscopically based on Gor'kov's theory \cite{zhu}.
This derivation is closely compared to the GL free energy for the
superconducting state composed of two degenerate components.
Applying the boundary conditions at $y=0$ and $y=W$ where $W$ is the 
width of the interface we derive the following current phase relation 
\cite{stefan4}
\begin{equation}
J(\phi)=\widetilde J_c \sin(\phi+\phi_c),
\end{equation}
where $\phi$ is defined as the relative phase difference between 
the two superconductors and $\phi_c$ is the intrinsic phase difference.
$\widetilde J_c$ is the Josephson critical current density.
For the type of junction that we consider where the insulator has 
a definite thickness and it is not a point
contact as in the case treated by Barash {\it et al.} \cite{barash}, 
the Josephson effect is strongly directional dependent and
the possibility for the tunneling of the Cooper pairs becomes maximum
when the trajectory of the Cooper pair is vertical to the interface.
So the total momentum of the order parameter functional depends only
on the orientation of the interface. This dependence enters the 
current phase relation via the $\widetilde J_c$, and $\phi_c$. 
We consider a corner Josephson junction between
the superconductor $A$ and the superconductor $B$ 
as seen in Fig. \ref{figdyn.fig}(b). 
The orientation of the $a$ and $b$ crystallographic 
axes are at right angles with the junction edges. 
We map the two segments of this junction each of length $l/2$ into a one dimensional axis. 
The characteristic phases $\phi_{c1}$, $\phi_{c2}$ for the 
two segments, for the 
pairing states that we consider can be seen in table \ref{tablephic}.
The superconducting phase difference $\phi$ across the junction is 
then the solution of the sine-Gordon equation
\begin{equation}
  \frac{d^2 { \phi}}{dx^2} - \frac{d^2 { \phi}}{dt^2} = 
J(\phi) + \gamma \frac{d{ \phi}}{dt}
,~~~\label{eq01} 
\end{equation}
where $\gamma$ is the damping constant which depends on the temperature. The inline boundary condition reads
\begin{equation}
\frac{d { \phi}}{dx}\left|_{x=0,l}\right. =\pm \frac{{
I}}{2},   
~~~\label{eq02} 
\end{equation}
where $I$ is the inline bias current.
The length $x$ is normalized in units of the 
Josephson penetration depth given by 
\begin{equation}
\lambda_J=\sqrt{\frac{\hbar c^2}{8\pi e d \widetilde J_c}}, 
\end{equation}
where $d$ is the sum of 
the penetration depths in two superconductors plus the thickness of the 
insulator layer. The time $t$ is in units of the inverse of the Josephson plasma frequency 
\begin{equation}
\omega_0^{-1}=\lambda_J/c_S.
\end{equation}
We can classify the different solutions obtained from Eq. \ref{eq01} 
with their magnetic flux content, in units of the flux quantum $\Phi_0$ 
\begin{equation}
\Phi=\frac{1}{2\pi}(\phi_R-\phi_L),
\end{equation}
where $\phi_{L(R)}$ is the value of the phase at the left(right) edge of the junction. 

\section{large junction limit} 

\begin{figure}
\centerline{\psfig{figure=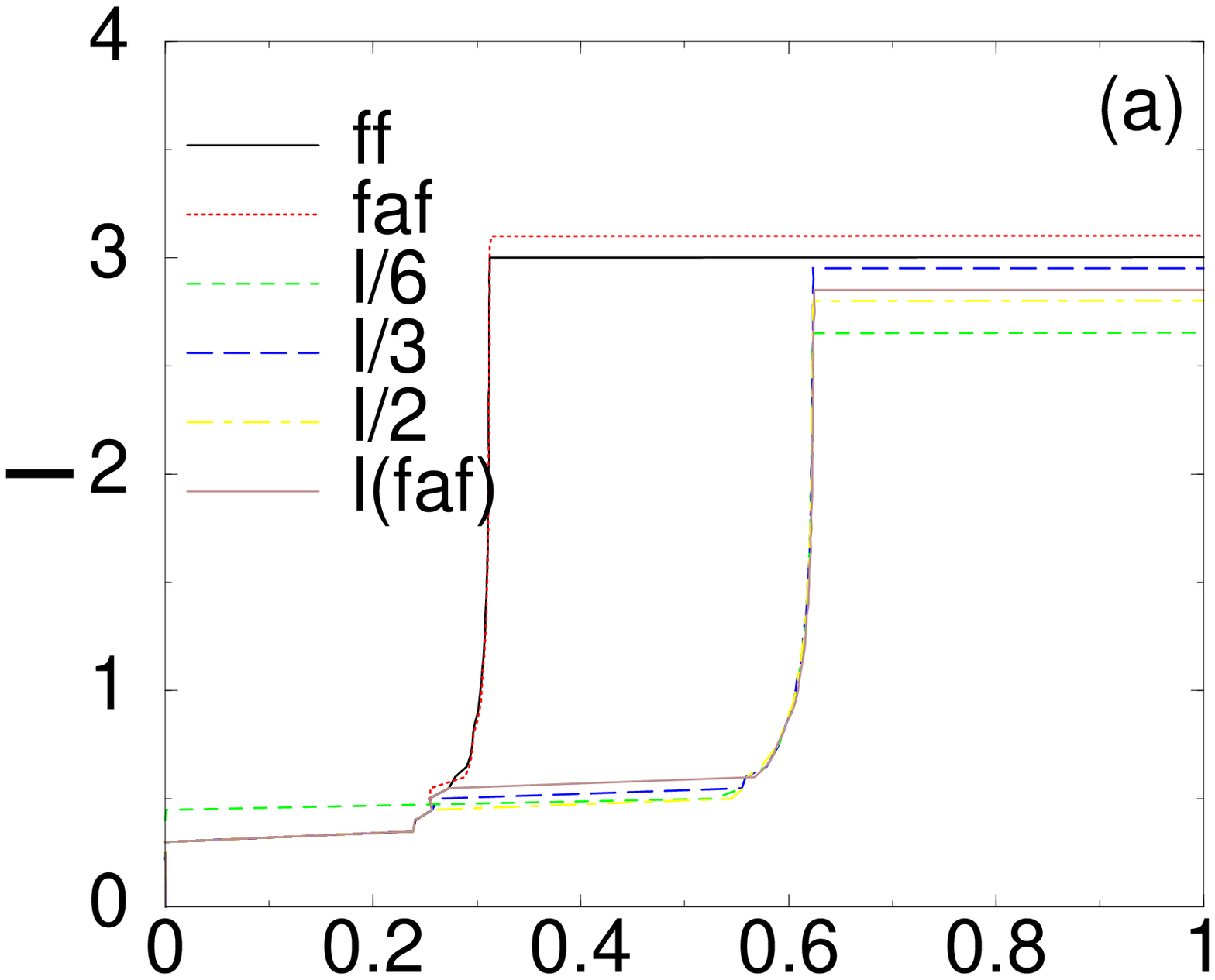,width=8.5cm,angle=0}}
\centerline{\psfig{figure=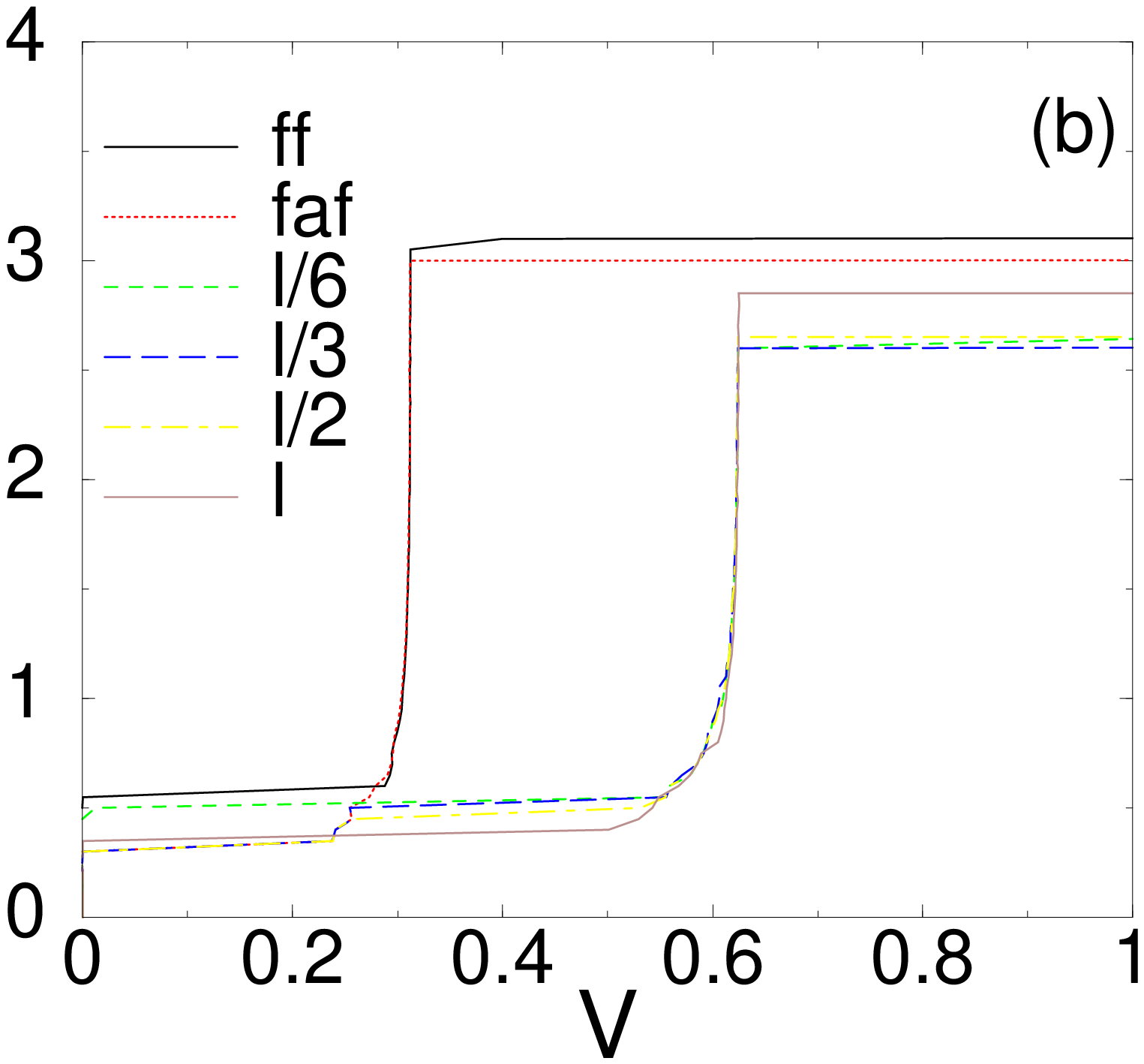,width=8.5cm,angle=0}}
\caption{$I-V$ characteristics for the inline geometry, for the first and 
second ZFS, for the $E_u$ pairing symmetry. 
The solutions for the first ZFS are the $ff$, $faf$ corresponding 
to a bound fluxon or antifluxon in the junction's center.
For the second ZFS the solutions are labeled by their relative distance $l/x$, 
where $l=20$ is the junctions length and $x=1,2,3,6$ respectively.
}
\label{IV.fig}
\end{figure}

A $4^{th}$ order Runge Kutta method with fixed time step $\Delta t=0.01$, 
was used for 
the integration of the equations of motion. 
The number of grid points is $N=1000$, and the junction length is 
$l=20$. 
The damping coefficient $\gamma=0.01$ is used in all the calculations.
The $I-V$ characteristics for the first and second ZFS 
(corresponding to one and two fluxons moving into the junction) 
are seen in Fig. \ref{IV.fig}(a) for the $E_u$ and in 
Fig. \ref{IV.fig}(b) for the $B_{1g}\times E_u$ pairing state.
For the first ZFS two different modes of fluxon propagation exist, 
corresponding to the presence of bound fractional fluxon or 
antifluxon respectively at the junction center. 
For the $E_u$ case, due to the difference in the flux content 
of the bound solutions, 
the  
$ff$ has smaller critical current than the $faf$.
This is opposite to the $B_{1g} \times E_u$ case.
In the $B_{1g}$ wave case the $I-V$ curves 
for the first ZFS, for the $ff$ and $faf$ mode 
have equal critical currents \cite{stefan3}.

\begin{figure}
\centerline{\psfig{figure=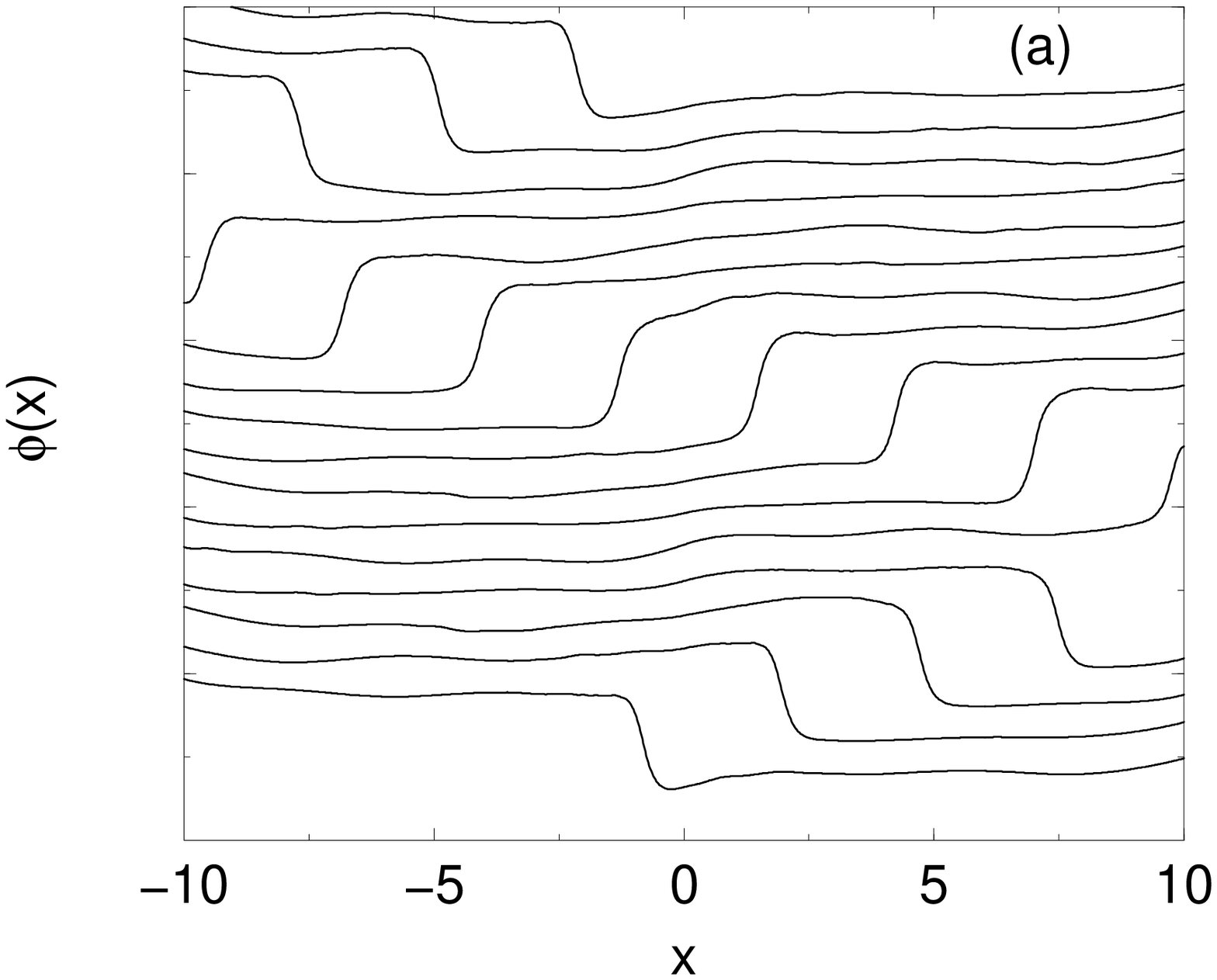,width=8.5cm,angle=0}}
\centerline{\psfig{figure=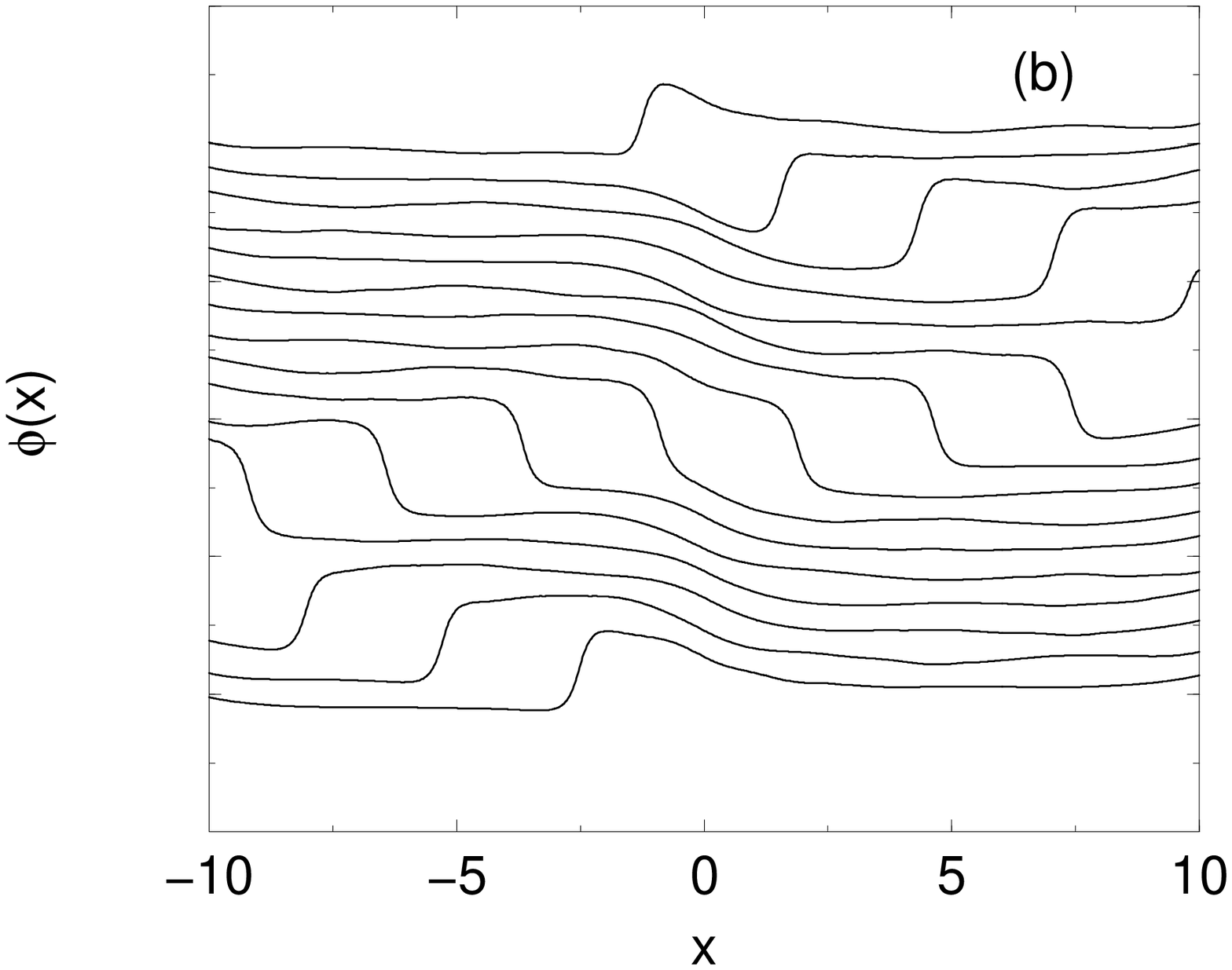,width=8.5cm,angle=0}}
\caption{Phase $\phi(x)$ vs $x$ 
for the solutions in the first ZFS,
at various instants, during one period 
separated by $\Delta \tau=2.8$.
The curves are shifted by $0.5$ to avoid overlapping.
$l=20$, $I=1.6$, $\gamma=0.01$: (a) $ff$,
(b) $faf$. The pairing symmetry is $E_u$.}
\label{1ZFS.fig}
\end{figure}

For the $E_u$ case, the external current cannot move 
the $faf$ with $\Phi=-0.75$ 
which is confined at $x=0$ (see Fig. \ref{1ZFS.fig}(a)). 
By applying the external 
current it emits an integer antifluxon ($AF$) which moves to the right  
and converts to an $ff$ with $\Phi=0.25$. The $AF$ hits the right boundary and transforms 
into a fluxon ($F$) which moves to the left. When the $F$ reaches the center
drags the $ff$ with it for a while and forms a fluxon with flux $\Phi=1.25$ which then 
breaks into a $ff$ and a $F$ moving to the left. 
The fluxon hits the left boundary transforms into an antifluxon 
which moves to the center where it meets the oscillating $ff$ and interacts 
with it forming a $faf$ and the period has been completed.
A full period of motion back and forth takes time $t=40$, and
since the overall phase advance is $4\pi$, in the relativistic limit
where $u=1$ reached at high currents, the dc voltage across the
junction will be $V=0.314$. 

We may also have the situation where the $ff$ with $\Phi=0.25$ 
exists at the junction center as seen in Fig. \ref{1ZFS.fig}(b). 
By applying an external current it emits an integer fluxon 
which moves to the left and converts to a $faf$ with $\Phi=-0.75$. 
The $F$ hits the left boundary and transforms to an $AF$ which moves 
to the right. When the $AF$ reaches the center it meets the oscillating $faf$ 
and forms a large antifluxon with $\Phi=-1.75$ which then splits into a $faf$ 
with $\Phi=-0.75$ and an integer antifluxon moving to the right. 
The antifluxon hits the right boundary, transforms into a fluxon which moves 
to the center where it meets the oscillating $faf$, 
interacts with it forming a $ff$ and the period has been completed. 
For the $B_{1g}\times E_u$ pairing state the $ff$ has magnetic flux 
$\Phi=0.75$ while the $faf$ has magnet flux $\Phi=-0.25$ and 
an integer fluxon or antifluxon propagates into the junction and 
interacts with the trapped fluxon or antifluxon. 

\begin{figure}
\centerline{\psfig{figure=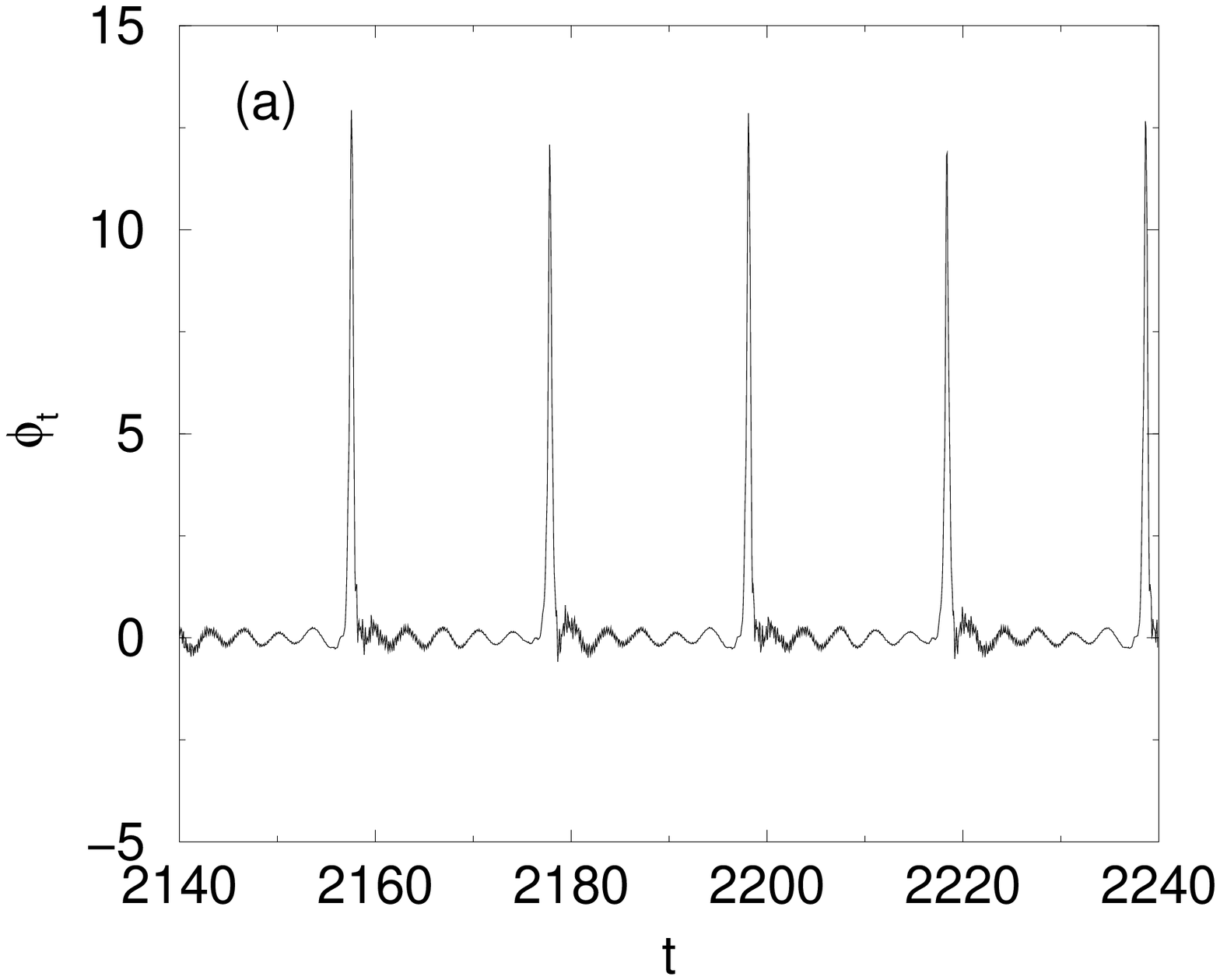,width=8.5cm,angle=0}}
\centerline{\psfig{figure=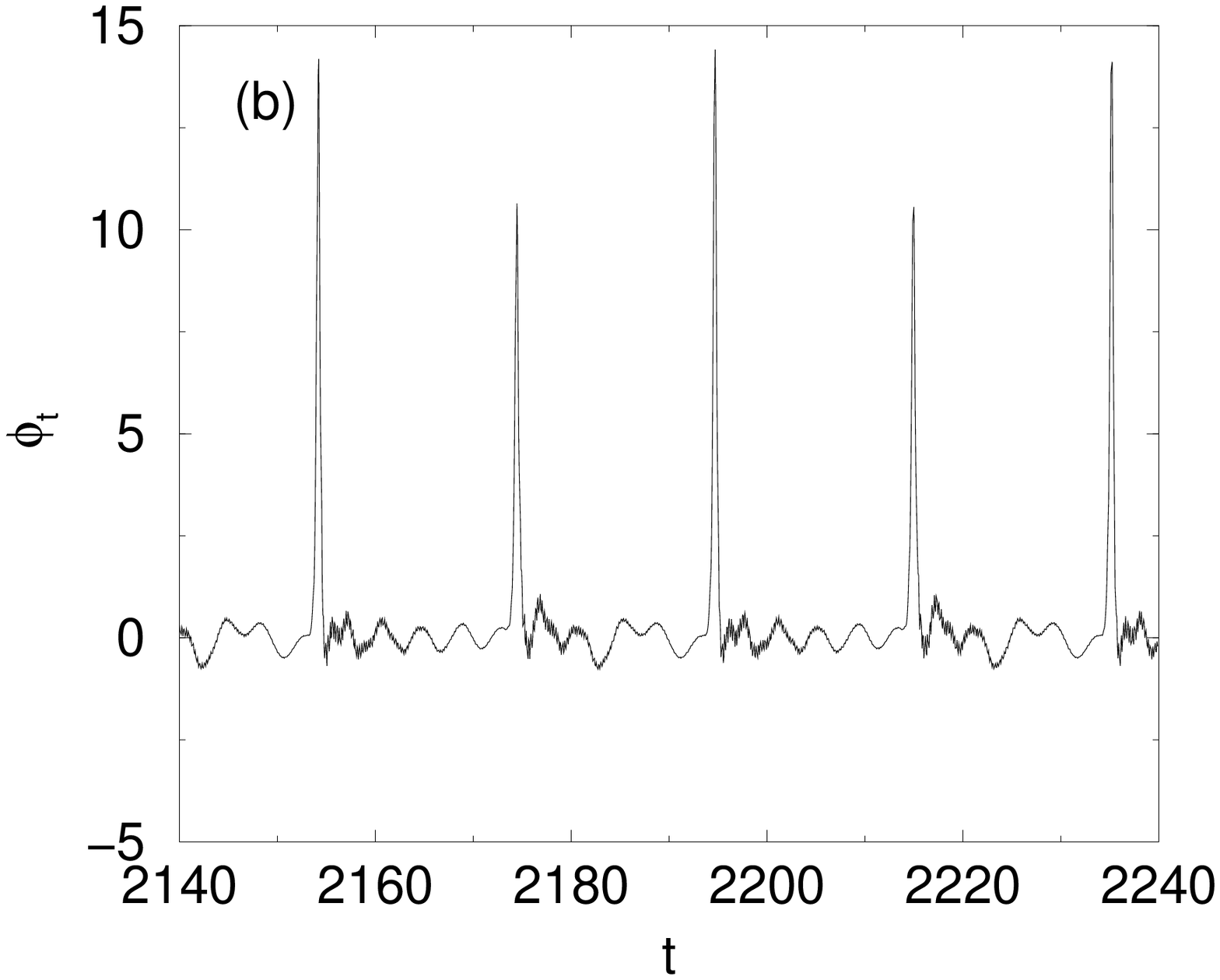,width=8.5cm,angle=0}}
\caption{Instantaneous voltage in the middle of the junction $(x=0)$ 
vs time $t$, for the solutions in the first ZFS. 
$l=20$, $\gamma=0.01$, $I=1.6$: (a) $ff$,
(b) $faf$. The pairing symmetry is $E_u$.}
\label{ft1ZFS.fig}
\end{figure}

The different character of the various fluxon solutions 
can also be seen from the plot of the instantaneous 
voltage $\phi_t$ at the center of the junction
for the various fluxon configurations. 
That plot is seen in Fig. \ref{ft1ZFS.fig} for the solutions 
regarding the first ZFS for the $E_u$ state.
During 
the time of one period three peaks appear in this plot by the time 
the fluxon (antifluxon) passes through the junction center. 
Note that the characteristic oscillations 
of $\phi_t$ between the peaks are due to the oscillation 
of the bound solution about the junction center. 
These oscillations become more distinguishable in the $faf$ case due 
its larger fluxon content. 
Note also the difference in height
between successive peaks in the $\phi_t$ vs $t$ diagram. 
This difference become more pronounced in the $faf$ case 
due to its larger flux content.
The plot of $\phi_t$ at the edges shows two peaks during the time  
of one period at time instants which differ by half 
a period. 
So for the first ZFS the $\phi_t$ vs $t$ plot 
can be used to 
probe the existence of $ff$ or $faf$ at the junction center.
For the $B_{1g}\times E_u$ pairing state (not presented in the figure)
the $ff$ has larger flux and the difference in the peak heights 
and the inter-peak oscillations become 
more distinguishable in the $ff$ case. 

\begin{figure}
\centerline{
\psfig{figure=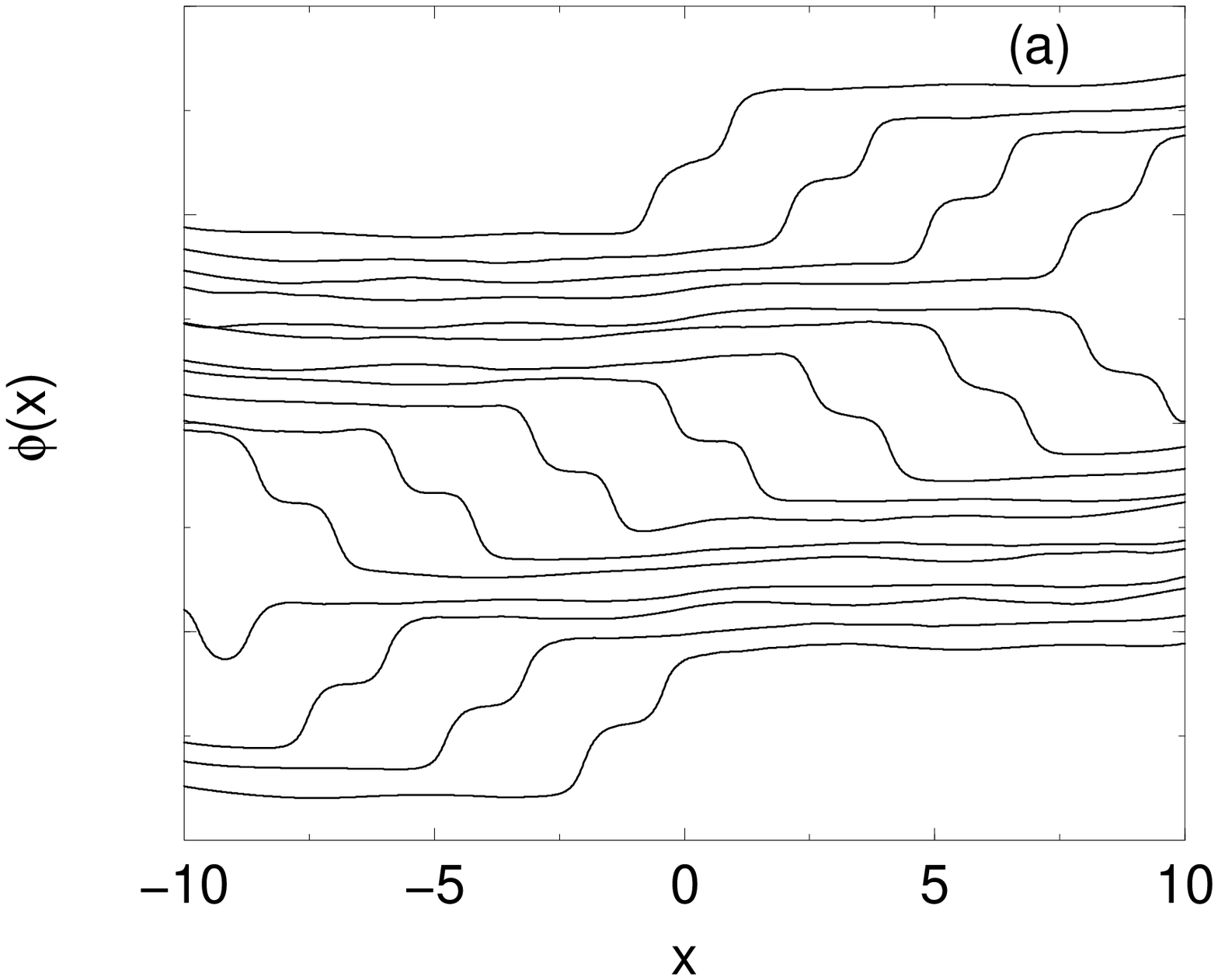,width=8.5cm,angle=0}
\psfig{figure=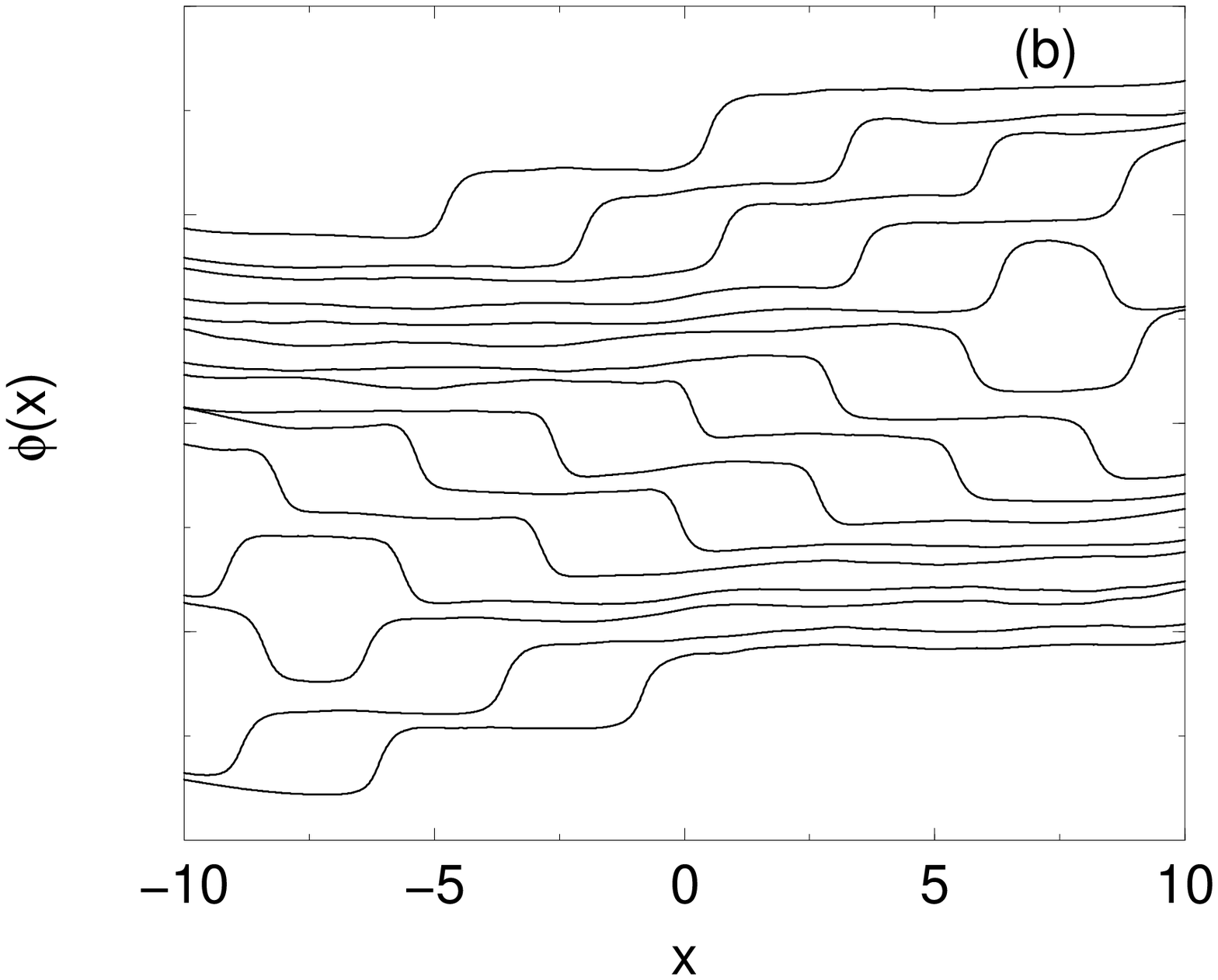,width=8.5cm,angle=0}}
\centerline{
\psfig{figure=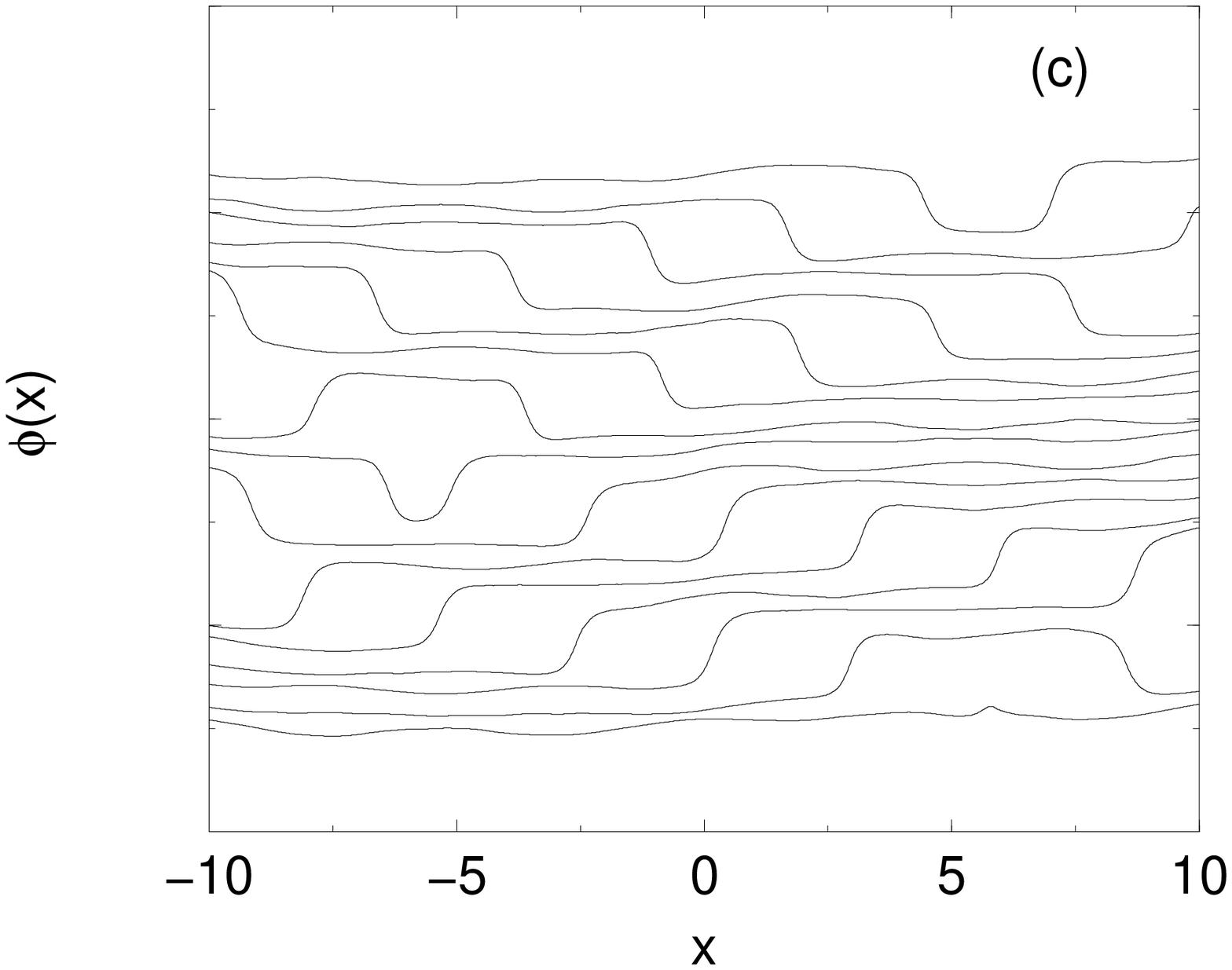,width=8.5cm,angle=0}
\psfig{figure=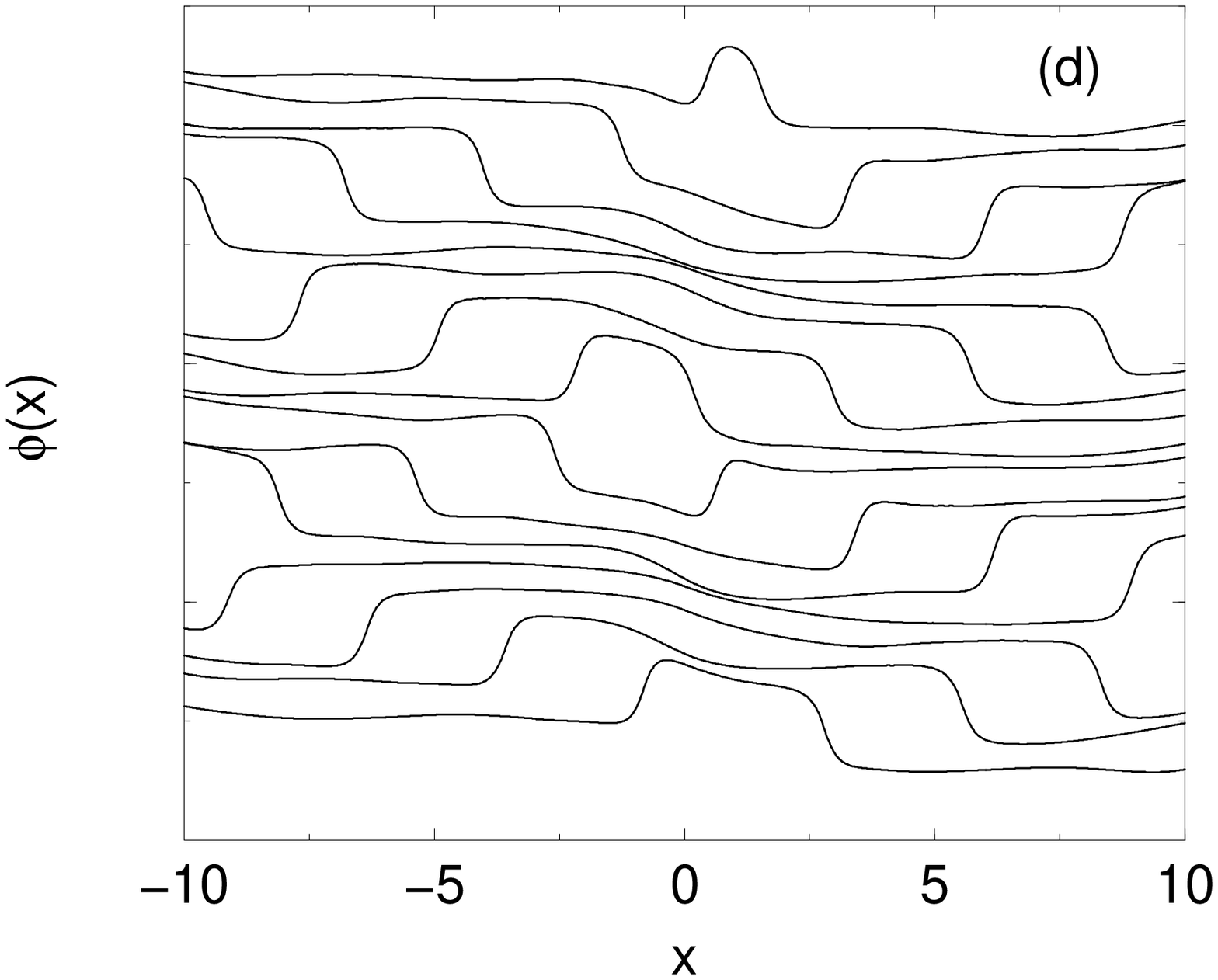,width=8.5cm,angle=0}}
\caption{Phase $\phi(x)$ vs $x$, 
for the solutions in the second ZFS, 
at various instants separated by 
$\Delta \tau=2.8$.
The curves are shifted by $0.5$ to avoid overlapping.
$l=20$, $\gamma=0.01$ :
(a) $l/6$, $I=1.6$, 
(b) $l/3$, $I=1.6$, 
(c) $l/2$, $I=1.6$,
(d)$l$, $I=1.6$. The pairing symmetry is $E_u$.}
\label{2ZFS.fig}
\end{figure}

For the second ZFS we found four two-fluxon configurations 
with distinct $I-V$ curves. Depending on the distance 
between the two vortices which is kept 
constant we can categorize the solutions as seen in Fig. \ref{2ZFS.fig}. 
So compared to the case of conventional $s$-wave superconductors \cite{lomdahl} 
junction we observe several curves for the second ZFS 
depending on the relative distance between the fluxons 
and this may be used to probe the presence of intrinsic magnetic 
flux. 

\section{shorter junction limit}
\begin{figure}
\centerline{\psfig{figure=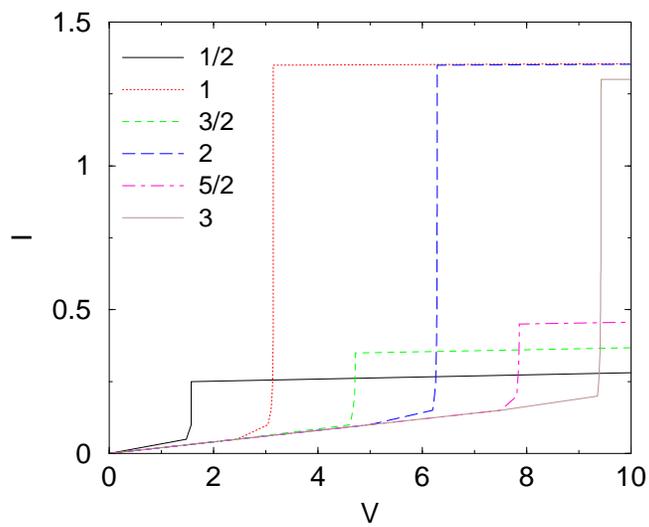,width=8.5cm,angle=0}}
\caption{$I-V$ characteristics for the inline geometry, 
for the $E_u$ pairing state, $l=2$. 
}
\label{IVpl=2.fig}
\end{figure}

\begin{figure}
\centerline{
\psfig{figure=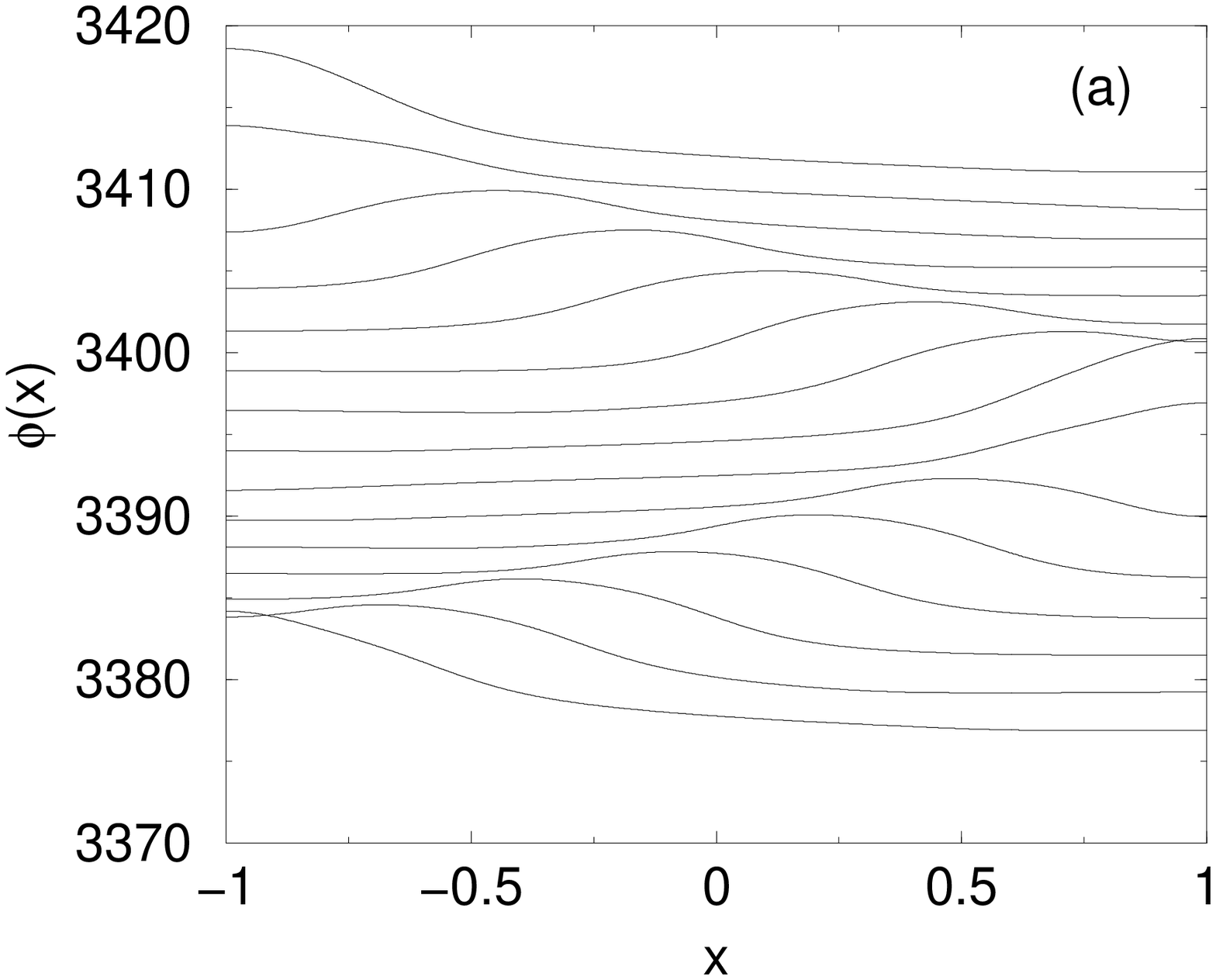,width=8.5cm,angle=0}
\psfig{figure=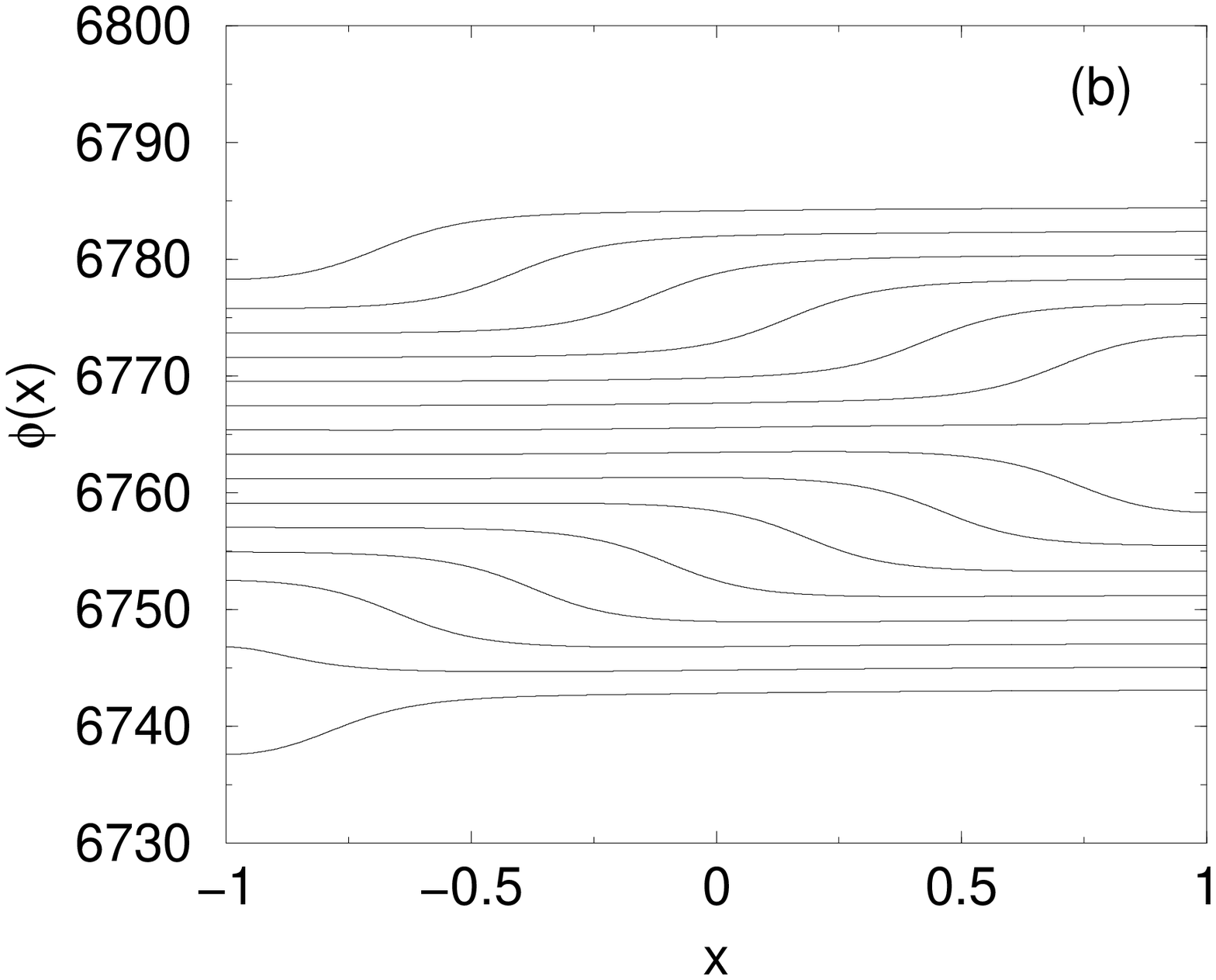,width=8.5cm,angle=0}}
\centerline{
\psfig{figure=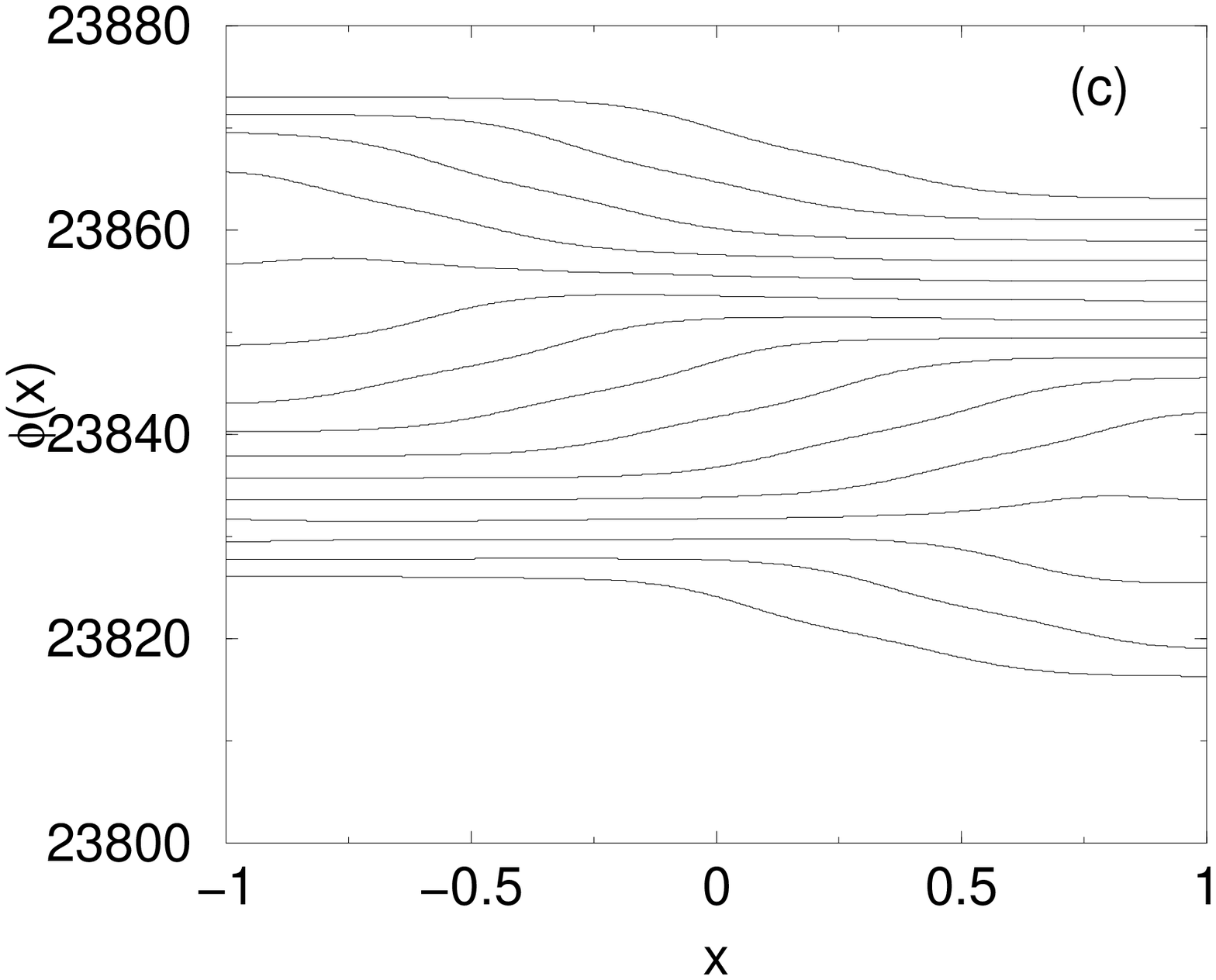,width=8.5cm,angle=0}
\psfig{figure=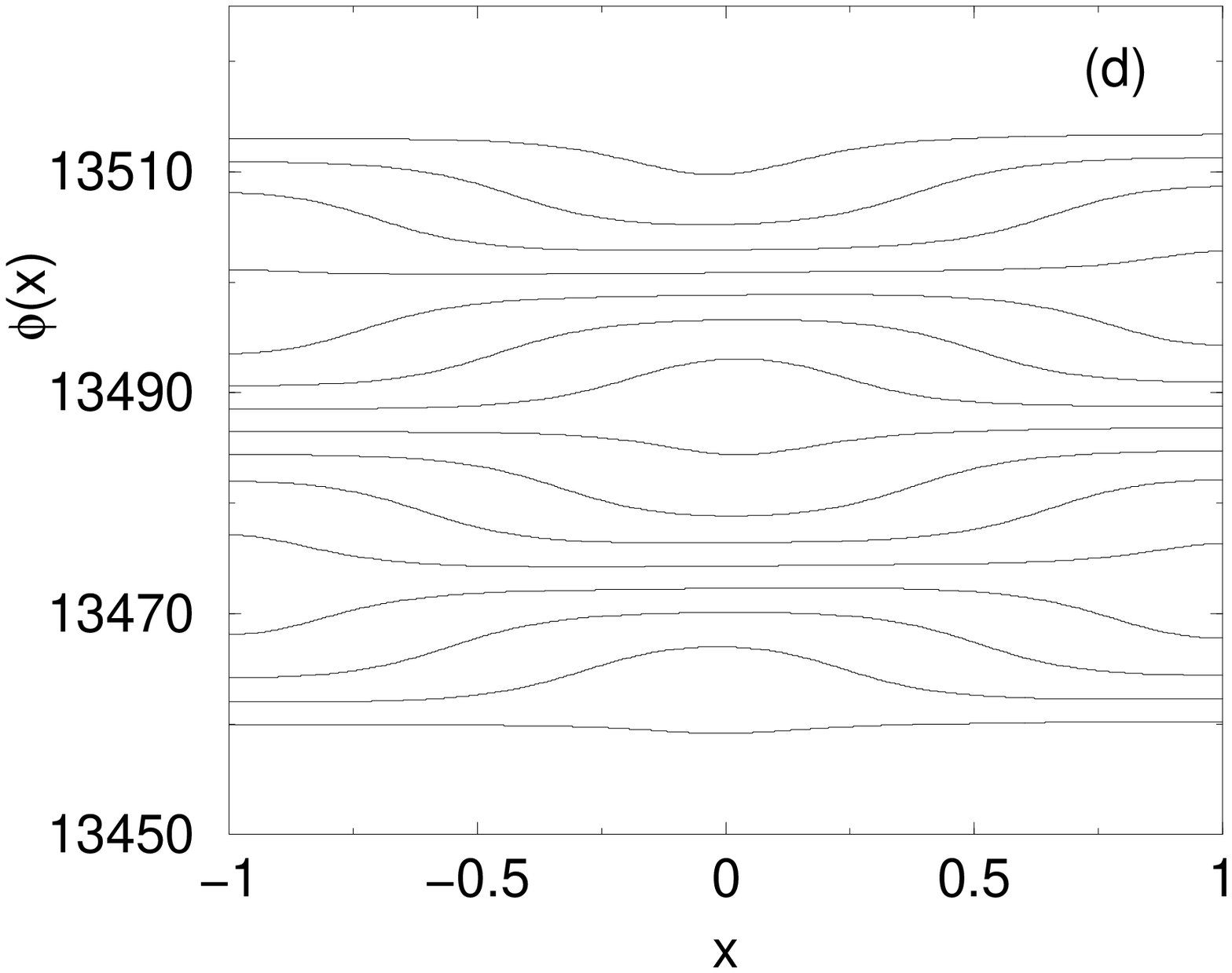,width=8.5cm,angle=0}}
\caption{Phase $\phi(x)$ vs $x$
for the solutions,
at various instants, during one period
separated by $\Delta \tau=0.2$.
The pairing state is $E_u$.
The curves are shifted by $0.5$ to avoid overlapping.
$l=2$, $I=0.25$, $\gamma=0.01$. }
\label{pl=2.fig}
\end{figure}

\begin{figure}
\centerline{
\psfig{figure=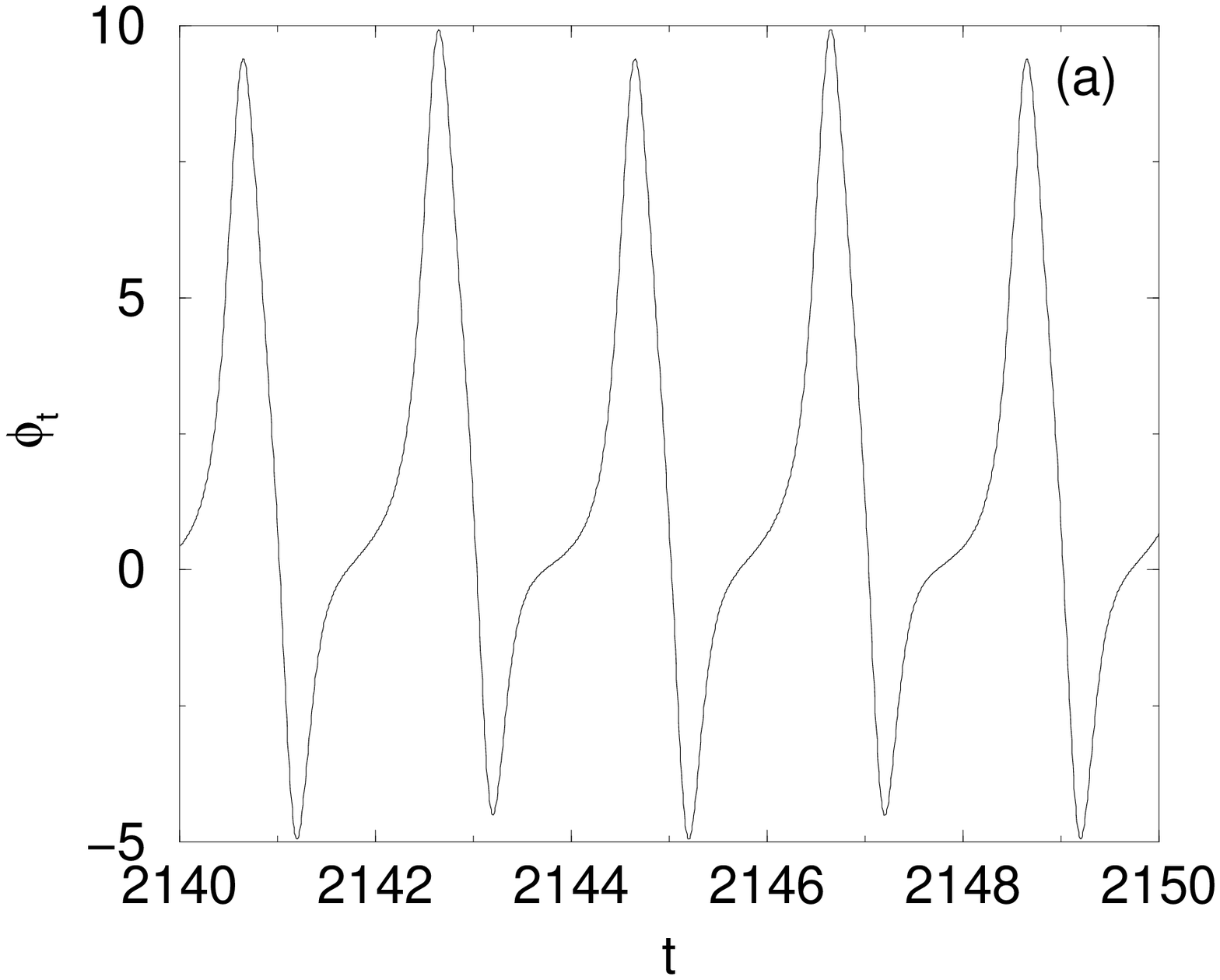,width=8.5cm,angle=0}
\psfig{figure=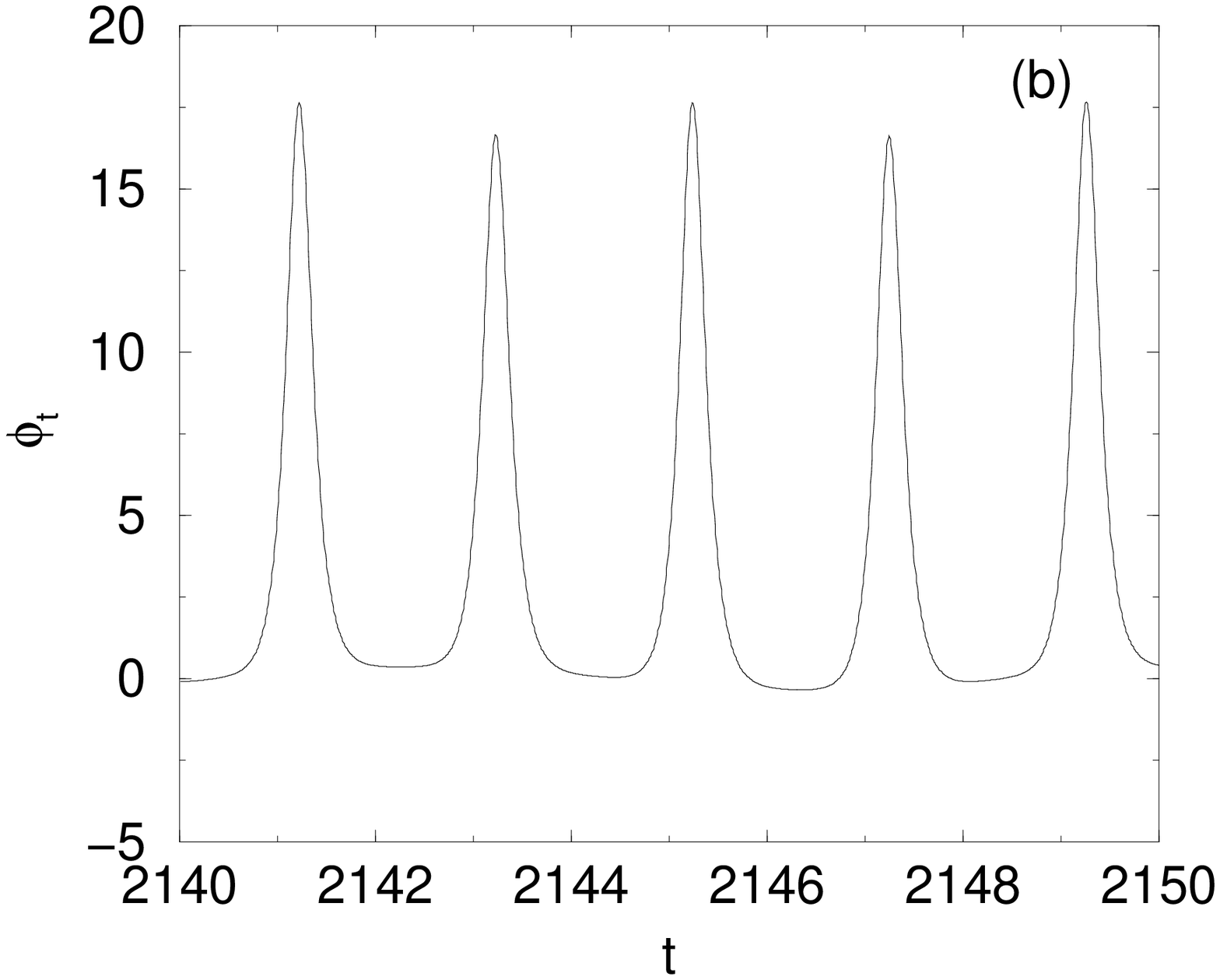,width=8.5cm,angle=0}}
\centerline{
\psfig{figure=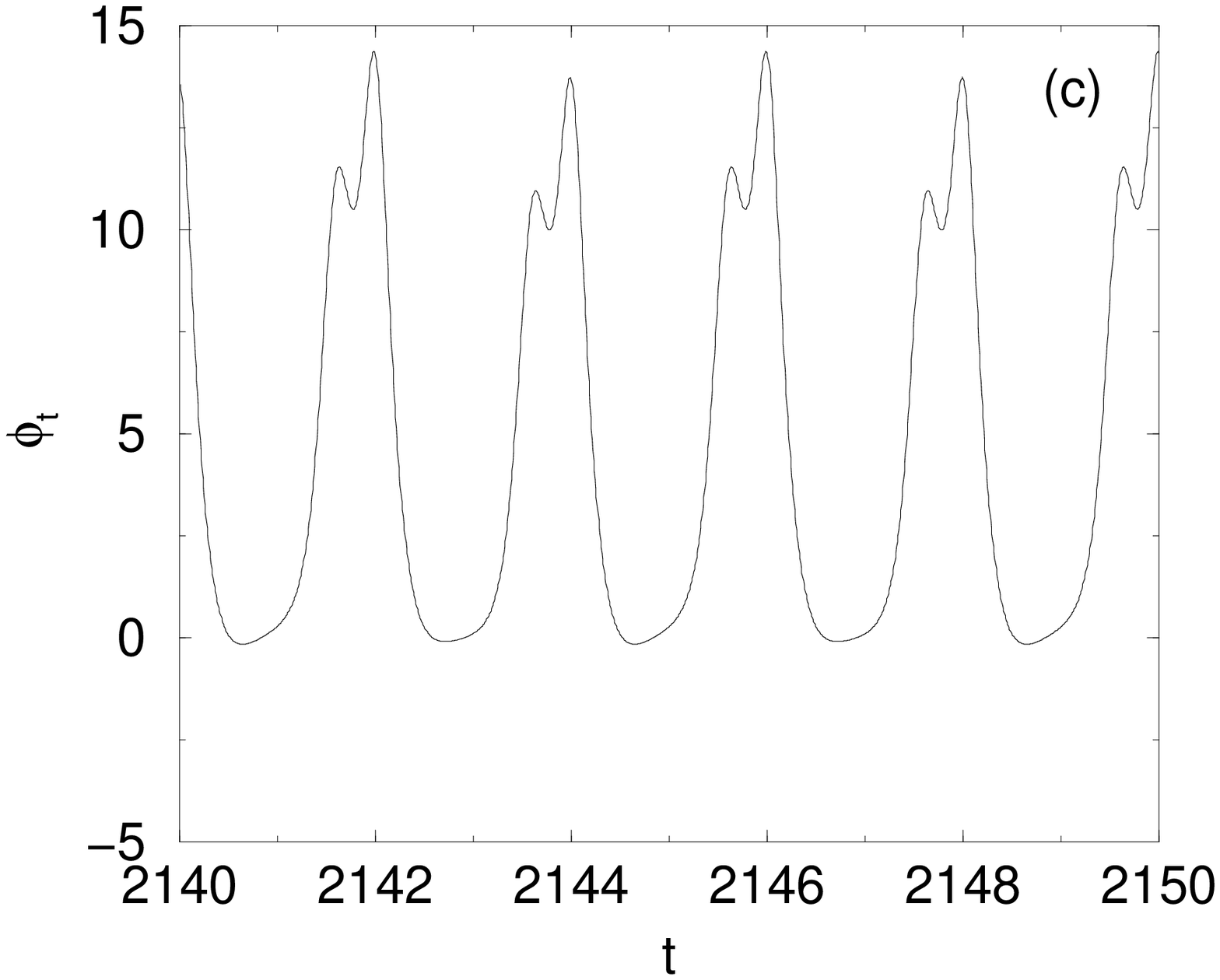,width=8.5cm,angle=0}
\psfig{figure=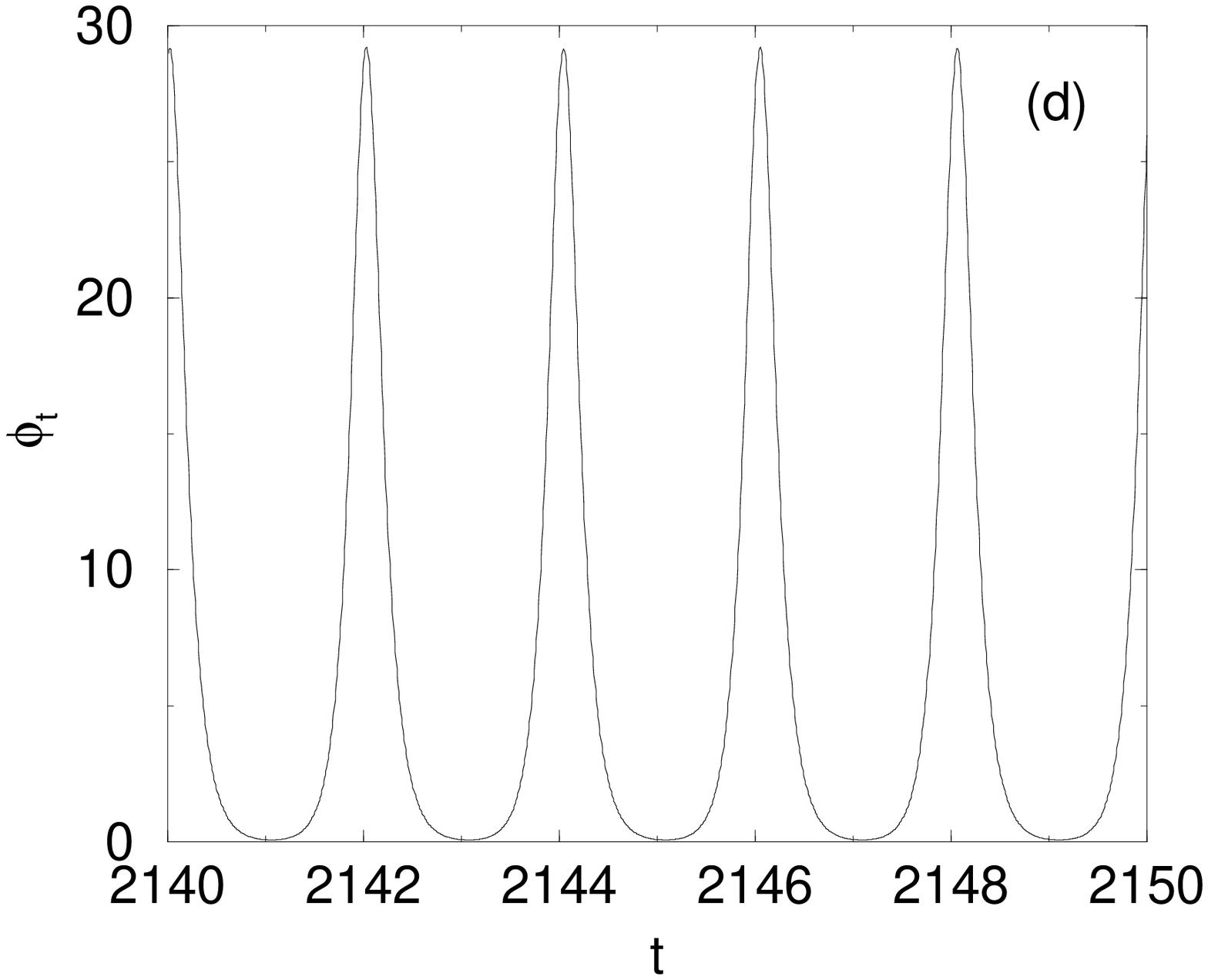,width=8.5cm,angle=0}}
\caption{
Instantaneous voltage in the middle of the junction $(x=0)$
vs time $t$, for the various solutions.
The pairing state is $E_u$.
$l=2$, $\gamma=0.01$, $I=0.25$.
}
\label{pl=2ft.fig}
\end{figure}

We now consider the case where the junction is of relative 
short length $l=2$.
We plot in Fig. \ref{IVpl=2.fig} the $I-V$ characteristics for the ZFS. 
We see that additional steps appear at half integer values of the 
dc voltage $V_1$
besides the ones that exist at integer values. 
In Fig. \ref{pl=2.fig}(a) we present $\phi(x)$, for the first ZFS,
for one period ($T=4$)
at various instances in time, which are separated by $\Delta \tau=0.2$.
For the $E_u$ case, for $t=0$ a $faf$ with $\Phi=-0.75$ 
exists at the junction center.
By applying the
external
current it converts to a $ff$ with $\Phi=0.25$ and a $faf$ 
with $\Phi=-0.75$. 
The $faf$ drags the $ff$ with it 
forming a combination of  fractional fluxon-antifluxon  ($ff-faf$), 
with negative magnetic flux $\Phi=-0.5$ which 
moves to the right.
It hits the right boundary and
the fractional antifluxon with $\Phi=-0.75$ goes to a 
fractional fluxon with $\Phi=1.25$, while the fractional fluxon with 
$\Phi=0.25$
goes to a fractional antifluxon with $\Phi=-0.75$, forming 
again a combination of 
fractional fluxon-antifluxon ($ff-faf$) with total flux $\Phi=0.5$, 
which moves to the left. 
Although the total flux is half integer the moving fluxon to the 
right direction has different structure than the combination 
that moves to the left and this explains the asymmetry 
between successive peaks in the 
$\phi_t$ diagram seen in Fig. \ref{pl=2ft.fig}(a).
The $ff-faf$ hits the left boundary transforms into an $ff-faf$ with 
$\Phi=-0.5$
which moves to the center where 
the period is completed.
A full period of motion back and forth takes time $t=4$, and 
since the overall phase advance is $2\pi$, in the relativistic limit 
where $u=1$ reached at high currents, the dc voltage across the 
junction will be $V=\pi/2$. This is indeed the value obtained from the 
numerical simulation as seen in Fig. \ref{IVpl=2.fig} for the 
solution labeled as $1/2$. 
Note that this value 
is half than the case where a full fluxon moves into the junction. 

In Fig. \ref{pl=2.fig}(c) we present $\phi(x)$, for the case where a 
bound combination of 
two fractional vortices with total magnetic flux $\Phi=1.5$ 
is propagating into the junction. 
In a junction of length $l=2$ 
the propagating fluxon accomplishes an overall phase advance 
of $6\pi$ in a full period $T=4$. 
Thus the dc voltage across the
junction will be $V=3\pi/2$. The ZFS seen 
in Fig. \ref{IVpl=2.fig} corresponds to a solution labeled as $3/2$. 
Note the the structure of the fluxon moving in the forward and 
backward direction is not the same and the successive 
peaks in the $\phi_t$ vs $t$ at the center of the junction 
do not have equal heights as seen in Fig. \ref{pl=2ft.fig}(c).

In Fig. \ref{pl=2.fig}(d) we present $\phi(x)$, for the case where 
the moving combination corresponds to magnetic flux equal to $2$. 
Thus the dc voltage across the
junction will be $V=2\pi$. The corresponding ZFS is labeled as 
$2$ in Fig. \ref{IVpl=2.fig}. Due to symmetry the $\phi_t$ vs $t$ 
at the center of
the junction will have the same height for the forward and backward 
direction as seen in Fig. \ref{pl=2ft.fig}(d).

For the $B_{1g}\times E_u$ state for the first ZFS, by increasing 
the bias current, the $faf$
with $\Phi=-0.25$ is transformed into a $ff$ with $\Phi=0.75$ 
and a $faf$ with $\Phi=-1.25$ forming a fractional fluxon with 
total flux $\Phi=-0.5$ 
that moves to the right. The reflected fluxons has $\Phi=0.5$ 
but differently to the $E_u$ case it is composed by a $ff$ with $\Phi=0.25$ 
and a $faf$ with $\Phi=-0.75$. This difference can be seen 
in the $\phi_t$ vs $t$ diagram at the center which is displaced in time 
by half a period compared to the $E_u$ case.
However the resulting 
$I-V$ are similar to the $E_u$ case.

\section{conclusions}

We analyzed the dynamics  of  fluxons moving in a frustrated 
Josephson junction with triplet pairing symmetry, 
and calculated the $I-V$ characteristics. 
The external current cannot move the $ff(faf)$ which is confined at $x=0$.
However the external current is able to reorient the $ff(faf)$ and emit
an integer fluxon(antifluxon).
For the first ZFS we found two distinct curves 
with different critical currents, 
which correspond to the case where the moving fluxon or antifluxon 
interacts with a bound fractional fluxon and antifluxon respectively. 
The critical currents are different for the $E_u$ and the 
$B_{1g}\times E_u$ pairing states and this can be used to distinguish the 
pairing symmetry.   
When there are more than one integer vortices moving in the 
junction, there is a possibility of different modes of fluxon 
propagation which correspond to different critical currents.

In the small junction limit,
due to the  presence of the internal flux, the moving
integer or half integer vortices will have internal structure
that is formed from the combinations of the static solutions.
The different modes in the $I-V$ diagram
exist both at integer and half integer values of the
dc voltage. However the $I-V$ is similar for the $E_u$ and 
$B_{1g}\times E_u$ and therefore it can not be used to 
distinguish between the two pairing states.

\section{acknowledgements}
Part of this work was done at the
Department of Physics, University of Crete, Greece.

\begin{table}
\caption{
We present the characteristic phases $\phi_{c1}, \phi_{c2}$ for the various
pairing symmetries.
$\phi_{c1}, \phi_{c2}$
is the extra phase difference in the two edges of the corner junction due to the
different orientations, of the $a$-axis of the crystal lattice.
}
\begin{tabular}{ccc}
        Pairing state & $\phi_{c1}$ & $\phi_{c2}$\\ \hline
        $E_u$ & $0$ & $-\pi/2$\\
        $B_{1g}\times E_u$ & $0$ & $\pi/2$
\end{tabular}
\label{tablephic}
\end{table}

\end{document}